\documentstyle[prb,aps,multicol,epsf]{revtex}
\tighten
\begin{document}
\draft
\vspace{2cm}
\title{
Superconducting fluctuations at low temperature}
\author{V. M. Galitski$^{1}$ and A. I. Larkin$^{1,2}$}

\address{$^1$Theoretical Physics Institute, University
of Minnesota, Minneapolis, MN 55455, USA\\
\vspace{0.5cm}
$^2$Landau Institute for Theoretical Physics,
Kossigin Str. 2, 117940, Moscow, Russia}

\maketitle
\begin{abstract}
  The effect of fluctuations on the transport and thermodynamic properties of
  two-dimensional superconductors in a magnetic field is studied 
  at low temperature $T \ll T_{c0}$.
  The fluctuation conductivity is calculated in the framework of 
  the perturbation theory
  with the help of usual diagram technique. It is shown that in the dirty case
  the Aslamazov-Larkin, Maki-Thomson and Density of States
  contributions are of the same order. At extremely low temperature
  $T/T_{c0} \ll \left( H -H_{c2}(0) \right) / H_{c2}(0)$ the total
  fluctuation correction to the normal conductivity is negative 
  in the dirty limit
  and depends on the external magnetic field logarithmically
  $\delta \sigma \propto \ln{\left( H - H_{c2}(0) \right)}.$ 
  In the non-local clean limit, the Aslamazov-Larkin contribution to conductivity
is evaluated with the aid  of the Helfand-Werthamer theory. 
The longitudinal and Hall  conductivities are found.
The fluctuating magnetization is calculated in the one-loop and two-loop
approximations.

\end{abstract}

\begin{multicols}{2}
\section{Introduction}

Over the last decade there has been continuing interest in Quantum Phase Transitions.
A particular attention has been focused on two-dimensional systems which
possess some unusual properties at low temperatures.
It is remarkable, that a phase transition at zero temperature  is possible in the framework
of the usual BCS theory of superconductivity. The transition temperature can be
suppressed
either by magnetic impurities or by magnetic field. It is interesting to find
the fluctuation conductivity as a function of the closeness to the transition in these cases.
The impurity-driven quantum phase transition has been considered by Ramazashvili and Coleman. \cite{Ram}
Their consideration was based on the renormalization group analysis of the Aslamazov-Larkin correction
to conductivity.
Fluctuations in an external magnetic field have been
considered in different systems and various limiting cases. \cite{AL,M,T,T2,Ami,AHL,BEL} However, up till now, there is no
consistent microscopic theory of superconducting fluctuations near $H_{c2}(0)$. The purpose of the 
present paper is to develop such a theory for two-dimensional superconductors in the dirty and clean limits.

We begin with a brief review of studies of fluctuations in superconductors. The subject was initiated
in the work of Aslamazov and Larkin. \cite{AL} The conductivity of fluctuating Cooper pairs was calculated
in zero magnetic field. Maki \cite{M} and Thomson \cite{T} included effects of electron scattering
off the fluctuations. It was found that there is another badly divergent contribution known as anomalous Maki-Thomson
correction. Physically, this correction is connected with the coherent scattering of the electrons by the impurities
and analogous to the weak localization correction.
The divergence can be removed by introducing a pair-breaking rate. Note, that experimental results at $T \sim T_{c0}$ can be
described by the Aslamazov-Larkin term only. This sugestes that the pair-breaking rate is relatively large in real superconductors.
Later, Thomson  and Maki returned to the issue and evaluated fluctuation correction  to the normal conductivity in finite fields.
Thomson \cite{T2} evaluated paraconductivity for small fields $T\sim T_{c0}$ and large fields parallel to a two-dimensional
superconducting sample. Ami and Maki \cite{Ami} considered a dirty three-dimensional superconductor put in an arbitrarily strong magnetic
field having calculated
the diagrams numerically. However, some technical simplifications that had been made in the paper (namely, the dynamic
fluctuations had been neglected) make the results inapplicable at very low temperature.
Moreover, three-dimensional case is very different from the two-dimensional one, as shown in the present paper.
Let us mention some relatively recent results in this field. In 1993 Aronov {\em et al.} \cite{AHL} developed a theory of transport
phenomena in the fluctuation region in the dirty, clean and superclean ($\omega_c \tau \sim 1$) limits. Their consideration
was based on the Ginzburg-Landau equations and, thus, is applicable for relatively small fields 
$H \ll H_{c2}(0)$ only. Beloborodov {\em et al.} \cite{BEL} have calculated the fluctuating conductivity 
of a three-dimensional granular superconductor in the region close to $H_{c2}(0)$.

Our paper is structured as follows. In Sec. \ref{sec:1} we consider a two-dimensional dirty sample
$T_{c0} \tau \ll 1$  (where $\tau$ is the scattering time). We calculate the total fluctuation correction
to conductivity which is described by the standard set of diagrams (see Fig. 1). We derive an analytical expression for
the fluctuation conductivity
in the region close to the transition line at low temperature
{\em i.e.} at $t = {T / T_{c0}} \ll 1$ and $h = \left( H - H_{c2}(T) \right) / H_{c2}(0) \ll 1$.
It is shown that in the case $t \gg h$ the total correction is positive and has the usual form
$\delta \sigma \propto T_{c0} \left( T - T_c(H) \right)^{-1}$, while at extremely low temperature $t \ll h$
(at zero temperature, in particular) the total correction becomes negative and logarithmically divergent
$\delta \sigma \propto  \ln{h}$.

In Sec. \ref{sec:3} we address the issue of fluctuations in clean superconductors. This problem is more complex,
since the elements upon which the diagrams are built (current vertices, cooperons {\em etc.}) are non-local
in the clean limit. We argue, that the corresponding operators can be found on the basis of the
Helfand-Werthamer theory. \cite{HW} We apply this theory to our problem and calculate all the necessary
values in the following limiting cases: $\omega_c \ll T$ or $\omega_c \tau \ll 1$
(where $\omega_c = e H_{c2}(0)/m \sim T_{c0} \left( T_{c0} / \varepsilon_F \right)$ is
the cyclotron frequency). This allows us to treat the magnetic field effects semiclassically.
The curving of the classical trajectories is taken into account by comparison with the Drude conductivity.
The longitudinal and Hall conductivities are found.
It is shown, that the fluctuation correction to conductivity in the clean limit is similar to the
one in the dirty limit, except for an additional cyclotron resonance-like pole of the second order which appears in the clean case.
At the end of Sec. \ref{sec:3}, we qualitatively discuss the effects of orbital quantization on
the fluctuation conductivity, {\em i.e.} Shubnikov-de Haas oscillations which become essential at
low temperatures $T \sim \omega_c$. 

In Sections \ref{sec:2} and \ref{tc}, we calculate thermodynamic properties of a superconductor. 
We find that magnetization is logarithmically divergent in the first approximation and exceeds Landau diamagnetism.
It is found, that in the clean case de Haas-van Alphen oscillations can become observable at high-enough temperature.
Under certain circumstances the oscillating part of the fluctuating magnetization represents the dominant effect.

In Sec. \ref{2l} we calculate the free energy and magnetization in the two-loop approximation for a dirty superconductor. 
We find that the divergence becomes more severe in the higher orders in the perturbation theory. We discuss 
the area of applicability of the results obtained.
We find that the fluctuation region is determined 
by $h \lesssim {\rm Gi}$, where ${\rm Gi}\, \sim \left(\varepsilon_F \tau \right)^{-1}$,
for low temperatures $t \ll h$ but it becomes wider $h \lesssim \sqrt{{\rm Gi\,} t}$ for relatively large temperatures $t \gg h$.

\section{Fluctuating conductivity}

\subsection{Dirty superconductors}
\label{sec:1}

The fluctuation correction to the conductivity beyond the Ginzburg region can be found in the perturbation theory.
There are terms of three different types describing the fluctuation conductivity in the first
(one-loop) approximation. The first one is the Aslamazov-Larkin (AL) term (see Fig. 1.1) which is
connected with the direct conductivity of the fluctuating Cooper pairs. The AL contribution to
conductivity is positive. Since some fluctuating pairs appear above the transition, the number of
normal electrons decreases. According to the Drude formula this leads to some decrease in the
conductivity of the normal electrons. This contribution is known as the Density of States (DOS)
term (see Figs. 1.5 and 1.8). It is clear, that this correction must be negative. The third term is
the Maki-Thomson (MT) contribution (see Fig. 1.2) which is connected with 
the coherent scattering of the normal electrons. The sign of the MT term is not prescribed. 

\begin{figure}
\epsfxsize=8.5cm
\centerline{\epsfbox{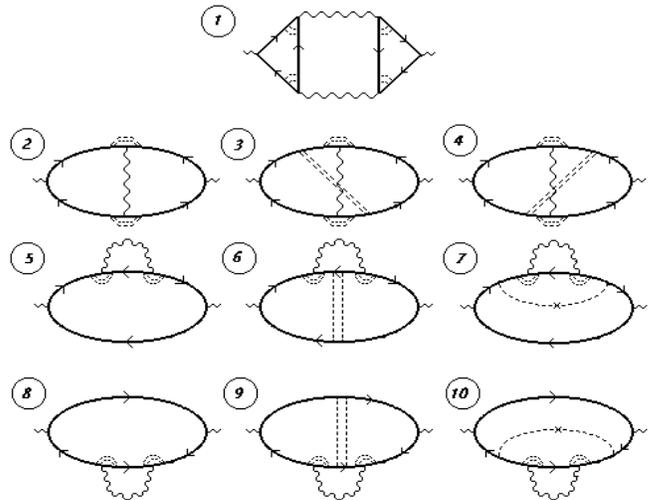}}
\caption{\label{fig:diag}
Impurity averaging diagrams contributing to conductivity in the first 
(one-loop) approximation.
}
\end{figure}

In the presence of impurities all these contributions must be averaged out over the
impurities positions. This can be done in the framework of the diagram technique developed long ago. \cite{AGD} 
There is a standard set of diagrams to be considered in our problem (see Fig. \ref{fig:diag}).

\narrowtext

These diagrams are built of the following elements:
A solid line represents the one-electron Green function which in zero field has the form (in the momentum representation):
\begin{equation}
\label{G0p}
{\cal G}_{\varepsilon}({\bf p}) = {1 \over i \varepsilon - \xi_{\bf p} + {\displaystyle i {\rm sgn \,} \varepsilon \over \displaystyle 2 \tau}},
\end{equation}
where $\varepsilon = (2 n +1) \pi T$ is the fermion Matsubara frequency and $\xi_{\bf p} = \varepsilon({\bf p}) - \varepsilon_F$
is the one-particle excitation spectrum. Here, we consider the quadratic spectrum.

In the presence of magnetic field ${\bf A}({\bf r})$ the Green functions change and contain the effects
of orbital quantization. However, in the presence of strong disorder $\omega_c \tau \ll 1$ or at relatively
high temperatures $T \gg \omega_c$, the discrete Landau levels are smeared out and the effects of magnetic field can be
treated semiclassically. This means that the Green function in the coordinate representation can be written as
\begin{equation}
\label{Gsem}
{\cal G}_{\varepsilon}({\bf r_1},\, {\bf r_2}) =
{\cal G}_{\varepsilon}^{(0)} ({\bf r_1} - {\bf r_2}) \exp{
\left( - i e \int\limits_{\bf r_1}^{\bf r_2} {\bf A} ({\bf s}) d{\bf s} \right)},
\end{equation}
where ${\cal G}_{\varepsilon}^{(0)}({\bf r})$ is 
the Green function in zero field and the path of integration in Eq.(\ref{Gsem}) is a straight line.
Let us note here, that the system of units $\hbar=c=k_B=1$ is used throughout the paper.
The magnetic field ${\bf H}$ is considered in the Landau gauge
${\bf A} = \left( 0,\, -Hx \right)$. 

Another element is the fluctuation propagator or interaction in the Cooper channel (wavy line).
It is a diagonal operator in the Landau representation. The corresponding matrix element has the form: \cite{AA}
\begin{eqnarray}
\label{K}
{\cal L}_n(\Omega) = {1 \over N(0)} 
\Biggl[ &\ln&{T \over T_{c0}}  \Biggr.
+ \psi\left({1 \over 2} 
+ {|\Omega| + \Omega_H \left( n + {1 \over 2} \right) \over 4 \pi T} \right)
\nonumber \\
&-& \psi\left({1 \over 2} \right) \Biggl. \Biggr]^{-1},
\end{eqnarray}
where $n$ corresponds to the $n$-th Landau level, $N(0)$ is the density of states per spin at the Fermi surface,
$\Omega = 2 \pi m T$ is the bosonic Matsubara frequency, corresponding to the total energy in the Cooper channel, 
$\Omega_H = 4 e {\cal D} H$ and  ${\cal D} = {1 \over 2} v_F^2 \tau$ is the diffusion coefficient. 
Note, that Eq.(\ref{K}) is obtained from the expression for the fluctuation propagator in zero-field
by the interchange of
${\cal D} q^2$ by $\Omega_H \left( n + 1/2 \right)$, with ${\bf q}$ being the total momentum in the Cooper channel. 

The shaded vertices in the diagrams are Cooperons, which describe the coherent scattering of two particles
off the impurities. The expression for this quantity has the following form: \cite{AA} 
\begin{equation}
\label{Coop}
C_n(\varepsilon_1,\, \varepsilon_2) =
{1 \over \tau} \, {\theta(-\varepsilon_1 \varepsilon_2) \over
\left| \varepsilon_1 - \varepsilon_2 \right| + \Omega_H \left( n + {1 \over 2} \right)},
\end{equation}
where  $\varepsilon_1$ and $\varepsilon_2$ are fermion Matsubara frequencies, corresponding to the electron
energies.

To calculate the total fluctuation correction to the DC conductivity we have to evaluate all the diagrams 1--10 as functions of
the external Matsubara frequency $\omega =2 \pi \nu T$, perform analytical continuation to the real frequency  axis, take the limit
$\omega \to 0$ and sum up all the contributions. The static term, corresponding to $\omega = 0$, is cancelled
out in the final result.

In the vicinity of $T_{c0}$ (transition temperature in zero-field) only the AL and anomalous  MT
terms are important. The typical arguments are as follows.
The point of superconducting transition is determined by the pole of the fluctuation propagator (wavy line). The AL
diagram contains two such lines. Thus, close to the transition the corresponding contribution is the most singular one.
Another singularity is due to the diffusion-like pole $\left( - i \omega + {\cal D} q^2 \right)^{-1}$ which appears in the MT term \cite{T}
(recall, that the MT process is connected with the coherent scattering  of electrons). At small ${\bf q}$ and
$\omega \to 0$ this yields a singular contribution.

Another simplification which can be made at $T \sim T_{c0}$ is the possibility to 
neglect dynamic fluctuations in the MT and DOS terms.
This means that instead of evaluating sum over the internal boson frequency $\Omega$ we can just take the first term
$\Omega=0$, which gives the most singular contribution. In the AL term the $\Omega$-dependence is considered
in the fluctuation propagators only and neglected in the current vertices.

The situation changes if a magnetic field is applied. \cite{Ami}
In this case, instead of integrating over ${\bf q}$,
we have to trace the corresponding operators over the Landau levels.
The AL diagram contains only one singular
fluctuation propagator ${\cal L}_0$ corresponding to the lowest Landau level,
since the current vertex is not a diagonal operator
in the Landau representation. Moreover, the small terms ${\cal D} q^2$, which
exist in zero-field, have to be
replaced by $\Omega_H \left(n + 1/2 \right) \sim T_{c0}$.
Obviously, the anomalous MT term does not possess any additional singularity
in this case. Thus, we conclude, that different diagrams should
give contributions of the same order if a large magnetic field is applied.

Let us now perform a representative calculation on the example of the AL term (see Fig 1.1).
The corresponding expression for the 
longitudinal component of the electromagnetic response tensor has the following form:
\begin{eqnarray}
&Q&_{1}(\omega) = - 4 e^2 c^2 \nu  \sum\limits_{n=0}^{\infty} \pi_{n \, n+1}^2 \nonumber\\
&\times& T\, \sum\limits_{\Omega}
 \biggl[ {\cal L}_n(\Omega) {\cal L}_{n+1}(\Omega - \omega) + {\cal L}_n(\Omega - \omega) {\cal L}_{n+1} (\Omega) \biggr] \nonumber\\
&\times& \left[
T \sum\limits_{\varepsilon} C_n(\varepsilon, \Omega-\varepsilon) C_{n+1}(\varepsilon - \omega, \Omega - \varepsilon) \right]^2,
\label{Qal}
\end{eqnarray}
where factor 4 is due to the spin, constant $c= 4 \pi N(0) {\cal D} \tau^2$ appears as a result of the integration
over $\xi$ in the local current vertex (see  (\ref{gamaloc})),
$\nu = e H / \pi$ is the number of states per unit area of a full Landau level and
$\pi_{n \, n+1} = \langle n \left| \left( -i \nabla + 2 e {\bf A({\bf r})} \right)_x
\right| n+1 \rangle  = \sqrt{ \left( n + 1 \right) e H}$ are matrix elements of the kinetic momentum.
$\omega$ is the Matsubara frequency corresponding to the frequency of the external electric field,
$\Omega$ and $\varepsilon$ are  the internal bosonic and fermionic Matsubara frequencies respectively. 

As we have already mentioned, the main singularity comes from the fluctuation propagator corresponding to the lowest Landau level.
Close to the transition it can be written as
\begin{equation}
\label{L0}
{\cal L}_0(\Omega) = {1 \over  N(0)} \, {1 \over h + 2 |\Omega| / \Omega_H},
\end{equation}
where $ h = \left( H - H_{c2}(T) \right) / H_{c2}(0)$.
Let us note, that $\Omega_H = 4 e {\cal D} H_{c2}(0) = {2 \pi \over \gamma} T_{c0}$ and the
bosonic frequency $\Omega$ is of the order of temperature. 
Thus, we conclude that at very low temperatures $t \ll h$ we can replace the sum over $\Omega$ in Eq.(\ref{Qal})
by an integral. At relatively high temperatures $t \gg h$ we can keep the first term in the sum only. If
$t \sim h$ we have to evaluate the sum. This also means that we have to consider the effects of dynamic fluctuations as well.

Let us discuss some simplifications that can be made in our case ($t \ll 1$). First of all, we can consider only
the first term $n=0$ in the sum over Landau levels in Eq.(\ref{Qal}). Only this term give a singular contribution
coming from ${\cal L}_0$. Next, we see that the sum over the internal
frequency in (\ref{Qal}) is determined by  $\Omega \sim T \ll \Omega_H$. This allows us to make expansions with
respect to $\Omega / \Omega_H \sim t$ everywhere except ${\cal L}_0$
with $\Omega$ being a Matsubara frequency, . With the same accuracy, we can replace the sum over the fermion energy $\varepsilon$
in Eq.(\ref{Qal}) by an integral.
%Let us note, that the corresponding calculations can be made without any simplifying assumptions. {\em I.e.} the analytical
%continuation on both frequencies can be made in general (for similar calculations see Ami and Maki [AM], Appendix B).
%The corresponding expressions for conductivity are cumbersome. It can be shown, that
%the result based on the simplifications described above coincides with the asymptotic form of the exact result.

Evaluating the integral over $\varepsilon$, 
we obtain from Eqs.(\ref{Coop}, \ref{Qal})
\begin{eqnarray}
&T& \sum\limits_{\varepsilon} C_0(\varepsilon, \Omega-\varepsilon) C_{1}(\varepsilon - \omega,
\Omega - \varepsilon) = \nonumber\\
&\phantom{.}& {1 \over 4 \pi \tau^2} \, 
\Biggl[ \Biggr. {1 \over \Omega_H - \omega}\, 
\ln{ \left( {3 \Omega_H / 2 + \left| \Omega - \omega \right| \over 
\Omega_H / 2 + \left| \Omega - \omega \right| + \left| \omega \right|} \right)}
\nonumber \\
&\phantom{.}& \phantom{ {1 \over 4 \pi \tau^2} \, \Biggl[ \Biggr.}
+ 
{1 \over \Omega_H + \omega}\, 
\ln{ \left( {3 \Omega_H / 2 + \left| \Omega \right| + \left| \omega \right| \over 
\Omega_H / 2 + \left| \Omega \right|} \right) } \Biggl. \Biggr].
\label{c01}
\end{eqnarray}
Now, we have to perform analytical continuation in the expression for the current response operator
(\ref{Qal}). In doing this, we can present the sum over the Matsubara frequency as an integral over the real frequency
with the function $\coth{\left( \Omega/ 2T \right)}$ 
which is chosen to generate poles at the points $2 \pi i m T$. \cite{Mah}
Making use of Eqs.(\ref{Coop} -- \ref{c01}) we obtain the following expression for the conductivity (within the logarithmic accuracy)
\begin{eqnarray}
\delta \sigma_1 = \lim_{i \omega \to 0} {Q_1(\omega \to i \omega) \over - i \omega} =
{e^2 \over \pi^2} \Biggl[ \alpha_1
\int\limits_0^{\Omega_{\rm max}} d \Omega \coth{\Omega \over 2 T} 
\Biggr. \nonumber \\
\Biggl. \times  {\Omega \over \Omega^2 + ({\Omega_H h \over 2})^2} 
+ \beta_1 \int\limits_0^{\infty} {d \Omega \over 2 T}
{1 \over \sinh^2{\Omega \over 2 T}}
{\Omega^2 \over \Omega^2 + ({\Omega_H h \over 2})^2} \Biggr],
\label{sigal}
\end{eqnarray}
where $\alpha_1 = 4/3$ and $\beta_1 = 2$ are just numbers.
One can see that the first integral in Eq.(\ref{sigal}) is
logarithmically divergent. This divergence appears as a result of
our expansions in $t$. Thus, it has to be cut off
at $\Omega_{\rm max} \sim T_{c0}$. The integrals in Eq.(\ref{sigal})
can be easily calculated. The result is
\begin{equation}
\delta \sigma =
{e^2 \over \pi^2}
\Biggl[ \alpha I_{\alpha}(h,t) + \beta I_{\beta}(h,t) \Biggr],
\label{Iab}
\end{equation}
with
\begin{equation}
\label{Ia}
I_{\alpha}(h,t) =
\ln{r \over h} - {1 \over 2 r} - \psi\left( r \right)
\end{equation}
and
\begin{equation}
\label{Ib}
I_{\beta}(h,t) = r \psi'\left( r \right) - {1 \over 2 r} - 1,
\end{equation}
where $r = {\displaystyle 1 \over \displaystyle 2 \gamma}\,
{\displaystyle h \over \displaystyle t}$ and $\gamma = 0.577$ is the Euler's
constant.

The other diagrams can be calculated analogously. The corresponding
contributions to conductivity
can be written in the same form as Eqs.(\ref{sigal}---\ref{Ib}).
Below we give the results in terms of constants $\alpha$ and $\beta$:

\begin{equation}
\begin{array}{ll}
\alpha_1 = {4 \over 3}, \phantom{abcd} & \beta_1=2; \\
\phantom{\alpha=1} & \phantom{\beta=2} \\
\alpha_2 = -2, \phantom{abcd} & \beta_2 = 2; \\
\phantom{\alpha=1} & \phantom{\beta=2} \\
\alpha_3 = \alpha_4 = -{2 \over 3}, \phantom{abcd} &  \beta_3 = \beta_4  = 0; \\
\phantom{\alpha=1} & \phantom{\beta=2} \\
\alpha_5 = \alpha_8 = - {3 \over 2}, \phantom{abcd} & \beta_5 = \beta_8 = - {3 \over 2}; \\
\phantom{\alpha=1} & \phantom{\beta=2} \\
\alpha_6 = \alpha_9 = {5 \over 3}, \phantom{abcd} & \beta_6 = \beta_9 =  {1 \over 3}; \\
\phantom{\alpha=1} & \phantom{\beta=2} \\
\alpha_7 = \alpha_{10} = {1 \over 2}, \phantom{abcd} & \beta_7 = \beta_{10} = {1 \over 2}; \\
\phantom{\alpha=1} & \phantom{\beta=2} \\
\alpha_{\rm tot} = -{2 \over 3}, \phantom{abcd} & \beta_{\rm tot} = {8  \over 3},
\end{array}
\end{equation}
where indexes correspond to a diagram number in Fig. 1 and $\alpha_{\rm tot}$,
$\beta_{\rm tot}$ describe the total correction to the conductivity, which can be written as:
\begin{equation}
\label{theres}
\delta \sigma =
{2 e^2 \over 3 \pi^2 \hbar}
\Biggl[ - \ln{r \over  h} - {3 \over 2 r} + \psi \left( r \right)
+ 4 \left( r \, \psi'  \left( r \right) - 1 \right) \Biggr].
\end{equation}

Let us consider some limiting cases.
If the temperature is relatively large $t \gg h$,
we obtain the following formula
for the fluctuation conductivity:
\begin{equation}
\label{lt}
\delta \sigma = {2 \gamma e^2 \over  \pi^2 \hbar}\, {t \over h}.
\end{equation}
If $H < H_{c2}(0)$, we can introduce $T_c(H)$ and rewrite Eq.(\ref{lt}) in the usual way
\begin{equation}
\label{lt2}
\delta \sigma = { 3 e^2 \over  2 \gamma \pi^2 \hbar}\,
{T_{c0} \over T - T_c(H)}.
\end{equation}

If $ H > H_{c2}(0)$, in the low-temperature limit $t \ll h$ we have
\begin{equation}
\label{st}
\delta \sigma = -{2 e^2 \over 3 \pi^2 \hbar}\, \ln{1 \over h}.
\end{equation}
One can see, that even at zero temperature a logarithmic singularity
remains and the corresponding correction is negative.

Let us note, that the fluctuating conductivity depends on the magnetic
field and temperature via their ratio $h / t$. The behavior of the
conductivity in the vicinity of the critical point $H = H_{c2}(0)$, $T=0$ depends on the 
way how one approaches this point. If the transition is driven by the magnetic field
and the temperature is zero, than the fluctuating correction is negative and logarithmically
divergent. If the magnetic field is fixed and $H \leq H_{c2}(0)$, than the correction
is positive and diverges as $\left( T - T_c(H) \right)^{-1}$. In the other cases, there is a crossover
between these two regimes. 

\narrowtext
\begin{figure}
\epsfxsize=8.5cm
\centerline{\epsfbox{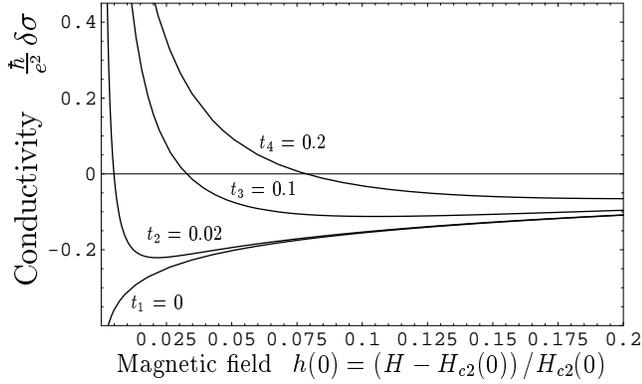}}
\caption{\label{fig:sigh}
Fluctuating conductivity (\protect\ref{theres}) as a function 
of magnetic field is plotted for four different temperatures. 
}
\end{figure}
\narrowtext
\begin{figure}
\epsfxsize=8.5cm
\centerline{\epsfbox{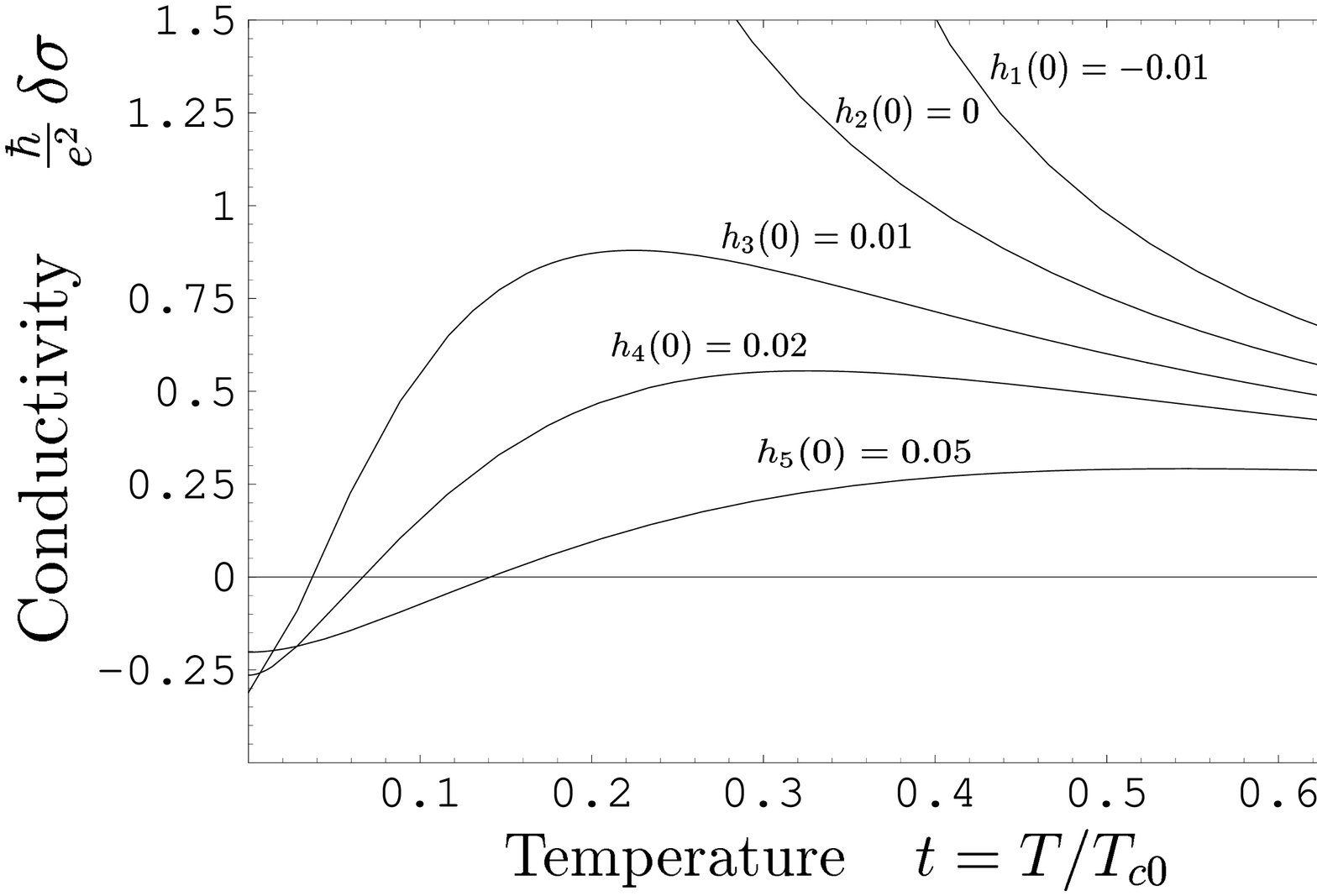}}
\caption{\label{fig:sigt}
Fluctuating conductivity (\protect\ref{theres}) as a function 
of temperature is plotted for five different magnetic fields. 
}
\end{figure}

The magnetic field dependence of the fluctuating conductivity is presented on Fig. 2. One can see,
that if the magnetic field is relatively large, than the total correction is negative. For any finite temperature,
there is a region close to $H_{c2}(0)$ where the correction is positive.

The temperature dependence of the fluctuating conductivity is shown on Fig. 3. It is interesting, that the conductivity is a non-monotonous
function of temperature if the magnetic field exceeds $H_{c2}(0)$.

\subsection{Clean limit}
\label{sec:3}

In this section, we investigate the fluctuation correction to the conductivity
in the limit $T_c \tau \gg 1$. 
In this case, usual expressions for the particle-particle bubble, fluctuation propagator and current 
vertices are inapplicable. To calculate the diagrams we have to find 
these quantities in the presence of the magnetic
field while taking into account their non-local structure. 
There are several effects associated
with the magnetic field applied. First of all, the superconducting 
transition itself is governed by the magnetic field at low temperature. 
Another effect is Shubnikov-de Haas oscillations in the conductivity due
to the quantization of the energy levels. However, if $\omega_c \tau \ll 1$ or $T \gg \omega_c$ 
the oscillating terms are exponentially small and can be neglected. 
Note, that
\begin{equation}
\label{oc}
\omega_c = {e H_{c2}(0) \over m} \sim T_{c0} \left( {T_{c0} \over \varepsilon_F} \right) \ll T_{c0}.
\end{equation}
In our formal derivation, we assume that either $\omega_c \tau \ll 1$ or
$\omega_c \ll T \ll T_{c0}$. This allows us to consider low temperatures 
without dealing with de Haas oscillations in the Green functions.
The effect of the orbital quantization on the fluctuation conductivity
will be briefly discussed at the end of this Section.
Moreover, there is a purely classical effect due to the
Lorentz force acting on the electrons forming fluctuating pairs. 
Namely, the magnetic field results
in curving of the classical trajectories. 
This curving leads to the cyclotron resonance and Hall effect in the fluctuation conductivity.
First, we consider fluctuations neglecting the curving, which is eligible if $\omega_c \tau \ll 1$.
Using the result obtained, we will be able to derive the formula valid in the superclean case $\omega_c \tau \sim 1$ 
as well.

We now proceed to calculate different blocks in the diagrams. Our calculation is based on the well-known
Helfand-Werthamer theory developed long ago. In the seminal paper, \cite{HW} Helfand and Werthamer evaluated the matrix element $C_0$ for the
Cooperon in a magnetic field, which determines the upper critical field $H_{c2}(T)$. They proved the following
mathematical statement which we will refer to as the Helfand-Werthamer (HW) theorem throughout the paper.

Let us consider an operator $\hat O$. 
Suppose, that its kernel in the coordinate representation has the following
form:
\begin{equation}
\label{ker}
O ({\bf r},\, {\bf r'}) = \tilde O ({\bf r -r'})\,
\exp{ \left( - 2 i e \int\limits_{\bf r}^{\bf r'}{\bf A}({\bf s}) d{\bf s} \right)}.
\end{equation}
Than, the operator can be written as
\begin{equation}
\label{hwt}
{\hat O} = \int  \tilde O ({\bf r})\, e^{-i {\bf r} \hat{\bbox \pi}} d^D r,
\end{equation}
where $\hat{\bbox \pi} = \left( \hat {\bf p} - 2 i e {\bf A}(\hat {\bf r}) \right)$ 
is the kinetic momentum, which can be expressed in terms of the
creation and annihilation operators in the Landau representation, 
and $D$ is the dimensionality of the system ($D=2$ in our case).

One can see that all the operators involved in our calculations satisfy the HW theorem. Namely, the particle-particle bubble
${\hat \Pi_{\varepsilon}(\Omega)}$, current vertex
$\hat\gamma_\alpha (\Omega,\omega)$ and the four Green function blocks $\hat B_{\alpha\, \beta} (\Omega,\omega)$ in
the coordinate representation can be written as a product of a function of the coordinate difference and the gauge factor.
In the temperature range under consideration, we can treat the magnetic field effects semiclassically which means that
the first factor $\tilde O$ in Eq.(\ref{ker}) can be considered in zero field.

To calculate the matrix elements of our interest we will do the following. First, we calculate an operator in zero-field
in the momentum representation $\tilde O ({\bf q})$. We apply the Fourier transformation to this function
and put the value obtained $\tilde O({\bf r})$ in Eq.(\ref{hwt}). Than, we evaluate the matrix
elements for this operator and perform 
the integration over ${\bf r}$. Finally, we perform the frequency summation left (over the fermion
energy $\varepsilon$).

Let us start with the calculation of the non-local fluctuation propagator which has the form:
\begin{equation}
\label{Lc}
\hat {\cal L} (\Omega)= {1 \over g^{-1} - \hat \Pi(\Omega)},
\end{equation}
where $g$ is the interaction constant and 
\begin{equation}
\label{tilP}
\tilde\Pi ({\bf q}, \Omega) =
T \sum\limits_{\varepsilon} \tilde\Pi_{\varepsilon} ({\bf q}, \Omega),
\end{equation}
with the particle-particle bubble $\tilde\Pi_{\varepsilon} ({\bf q}, \Omega)$ defined by
Eq.(\ref{bubble}). Note, that in the clean limit we can neglect the impurity dependence in 
$\hat \Pi_{\varepsilon}(\Omega)$ and in the fluctuation propagator.

The matrix elements can be calculated by expressing $\hat {\bbox \pi}$ in terms
of  the creation and annihilation operators $\hat{a}^{\dagger}$ and $\hat{a}$ and expanding the exponentials. \cite{Amb} 
One obtains
\begin{equation}
\label{Amb}
\exp{ \left( - i {\bf r} \hat{\bbox \pi} \right)} =
e^{-{\rho^2 \over 2}} 
\sum\limits_{k,\,l=0}^{+ \infty}
{\left( -i \rho \right)^{k+l} \over k!\, l!} \left(\hat a^{\dagger}\right)^k
\hat a^l
e^{-i \phi \left( l - k \right)},
\end{equation}
where $\rho= r / \sqrt{2} r_H$ and $r_H = \sqrt{2 e H}$ is the magnetic length. 
Due to the integration over $\phi$ only the diagonal matrix elements
survive and we have the following expression 
\begin{equation}
\label{Pn}
\Pi_n (\Omega) = (-1)^n r_H^2  T \sum\limits_\varepsilon \int\limits_0^{\infty}
d{q^2} \tilde\Pi_{\varepsilon} (q, \Omega) e^{-q^2 r_H^2} L_n (2 q^2 r_H^2),
\end{equation}    
where $L_n$ is the Laguerre polynomial of the $n$-th order.
%Let s repeat, that this expression has been derived long ago 
%by Ambegaokar {\em et al.}

At low temperature, we can replace the sum over $\varepsilon$ by 
an integral and we have
\begin{eqnarray}
&\Pi_n&(\Omega) = N(0) 
\Biggl[ \ln{ \left( 2 \Lambda \right) } \Biggr. \nonumber \\ 
\Biggl. &\phantom{\Pi}& - (-1)^n \int\limits_0^{\infty} dx
\ln{ \left( \lambda + \sqrt{\lambda^2 + x} \right) } e^{-x} L_n(2x)
\Biggr],
\label{PP}
\end{eqnarray}
where we have introduced $\lambda = {|\Omega| r_H \over v_F}$ being the lower limit of integration over
$\varepsilon$ and $\Lambda = {2 r_H \omega_D \over v_F}$,
which is the BCS high-energy cut-off. Obviously,
$\lambda \sim {T / T_{c0}} \ll 1$ and $\Lambda \gg 1$.

Let us realize, that to find the most singular contribution to the conductivity we need to know
$\Pi_0(\Omega)$ and $\Pi_1(\Omega)$ only. Making expansions with respect to $\lambda$ in Eq.(\ref{PP}),
one obtains
\begin{equation}
\label{pi0}
\Pi_0(\Omega) = N(0) \left[ \ln{\left( 2 \sqrt{\gamma} \Lambda \right)} - \sqrt{\pi}\lambda \right]
\end{equation}
and
\begin{equation}
\label{pi1}
\Pi_1(\Omega) = N(0) \left[ 1 + \ln{\left( 2 \sqrt{\gamma} \Lambda \right)} + \lambda^2 \right].
\end{equation}

Thus, the fluctuation propagator, corresponding to the lowest Landau level
can be written in the vicinity of the transition as follows
\begin{equation}
\label{l0c}
{\cal L}_0(\Omega) = {2 \over N(0)} {1 \over h + \sqrt{\gamma \over \pi}\, |\Omega| / T_{c0}},
\end{equation}
where $T_{c0}$ is the transition temperature in zero field, $h = \left( H - H_{c2}(T) \right) / H_{c2}(0)$
and $e H_{c2}(0) = {2 \pi^2 \over \gamma} \left( {T_{c0} \over v_F} \right)^2$ is the upper critical field
at zero temperature.    

The current vertex can be evaluated in the same fashion as 
the fluctuation propagator.  However, the corresponding
calculation is more cumbersome (see Appendix). 
In the momentum representation the vertex has the form: 
\begin{equation}
\label{gama1}
{\bbox \gamma} ({\bf q}; \Omega, \omega, {\varepsilon})
= {2 \pi N(0) \over \omega} \, {{\bf q} \over q^2} \,  
\sum\limits_{i=1}^4 \eta_i(\varepsilon,\Omega,\omega) f(\epsilon_i,q) \, ,
\end{equation}
The corresponding functions and constants are defined in the Appendix by formulae (\ref{ugol}---\ref{e23}). 
Eq.(\ref{gama1}) is large mostly
due to the theta-functions. However its q-dependent part has a simple form 
${{\bf q} \over q^2} f(q)$, where $f(0) = 0$ and $f(\infty)$ is finite.

Making use of the Helfand-Werthamer theorem, we obtain the 
following expression for the operator ${\bbox \gamma}$:
\begin{eqnarray}
\label{g2}
{\hat{\bbox \gamma}} (\Omega, \omega,\varepsilon) &=& 
2 \pi i N(0)  \sum\limits_{i=1}^4 {\eta_i(\varepsilon,\Omega,\omega) \over \omega} 
\int\limits_0^{\infty} dq f(\epsilon_i, q) 
\nonumber \\ 
&\phantom{=}& \times 
\int\limits_0^{\infty} dr\, r J_1(qr)
 \int\limits_0^{2 \pi} {d \phi \over 2 \pi} 
{{\bf r} \over r} \exp{\left( - i {\bf r}\hat{\bbox \pi} \right)}
\end{eqnarray}
Evaluating matrix elements and performing integration 
over $\phi$ and $r$, one can see, that
the current vertex possesses only near-diagonal 
non-zero matrix elements and they have the form:
\begin{eqnarray}
\label{gnn1}
&\phantom{.}&
\langle n \,| \hat\gamma^x (\Omega,\omega,\varepsilon) |\, n+1 \rangle =
\sqrt{2 \over n+1}  \pi N(0) r_H^3 \, (-1)^n 
\nonumber \\ &\phantom{.}& \times
\sum\limits_{i=1}^4 {\eta_i (\varepsilon,\Omega,\omega) \over \omega} 
\int\limits_0^{\infty}
d{q^2} f(\epsilon_i, q) e^{- r_H^2 q^2} L_n^{(1)}(2 r_H^2 q^2).
\end{eqnarray}

In order to calculate the most singular contribution to the conductivity
we have to know $\gamma_{01}$ only. Taking the corresponding integral
by parts one obtains
\begin{eqnarray}
\label{g01}
\gamma_{01} (\Omega,\omega,\varepsilon) &=& 
\sqrt{2} \pi N(0) r_H \sum\limits_{i=1}^4 {\eta_i (\epsilon,\Omega,\omega) \over \omega} 
\nonumber \\
&\phantom{=}& \times
\int\limits_0^{\infty} dq
{\partial f(\epsilon_i,q) \over \partial q} e^{ - r_H^2 q^2}.
\end{eqnarray}
To evaluate the remaining integrals we have to use the explicit expressions
for the functions $f(\epsilon, q)$ and $\eta_i(\varepsilon, \Omega, \omega)$ (see Appendix).
Using formula (\ref{ugol}), one obtains after elementary integration over $q$
\begin{eqnarray}
\label{g01d}
\gamma_{01} (\Omega,\omega,\varepsilon) &=& 
\sqrt{2} \pi N(0) r_H 
\sum\limits_{i=1}^4 {\eta_i(\epsilon,\Omega,\omega) \over \omega}
\nonumber \\
&\phantom{=}& \times \Bigl[  1 - \sqrt{\pi}\, 
|\delta_i| \, e^{\delta_i^2} {\rm erfc\,} |\delta_i| \Bigr],
\end{eqnarray}
where we have introduced $\delta_i = {\epsilon_i r_H / v_F}$. Now, we have to
perform the summation over the fermion frequency $\varepsilon$. For
$T \ll T_{c0}$ we can replace this sum by an integral over $\delta$ taken
in the  appropriate limits well-defined by the $\theta$-functions in 
Eqs.(\ref{gama}---\ref{e23}). The
corresponding indefinite integral can be easily evaluated
and we finally obtain
\begin{eqnarray}
\label{fingam}
 &\gamma_{01}& (\Omega,\omega) = {1 \over 2 \pi} \int d \varepsilon\, \gamma_{01}
(\Omega,\omega,\varepsilon)
\nonumber \\
&=& {\sqrt{2 \pi} N(0) v_F \over 8}
\sum\limits_{i=0}^4 {\eta_i \over \omega}
\, {\rm sgn\,} \delta_i \,
e^{\delta_i^2} {\rm erfc\,}
|\delta_i| \Biggl. \Bigr|{\phantom{.}}_{\rm limits}\,\, ,
\end{eqnarray}
where the limits of integration are determined by the theta-functions in $\eta_i$. 
Let us note, that at $\delta=\infty$ the expression written above vanishes,
while the other limits of integration are such that $\delta \sim T/T_{c0}
\ll 1$.

There are four terms in Eq.(\ref{gama}). The last two contain
the factor $\omega^{-1} {\rm sgn{\varepsilon}\,}$, while the first two are proportional to the
following factor
$$
{\tau \over 1 + \omega \tau\, {\rm sgn\,}{\varepsilon} } =
{{\rm sgn\,}{\varepsilon} \over \omega}
\left[ 1 -
{1 \over 1 + \omega \tau\, {\rm sgn\,}{\varepsilon} }
\right].
$$
Using Eq.(\ref{fingam}), one can see, that 
the $\omega^{-1}$--terms are cancelled out exactly.
Thus, the current vertex can be written (we keep only linear terms with respect
to the frequencies): 
\begin{eqnarray}
&\gamma_{01}& (\Omega,\omega) =
-{N(0) r_H \over \sqrt{2}}
\, {1 \over 1 + |\omega| \tau} \nonumber \\
&\times&	 \left[ 1 - {\sqrt{\pi} \over 2}
{r_H \over v_F} \left(
|\Omega| + |\Omega-\omega| + |\omega| \right) + o(t) \right].
\label{vot}
\end{eqnarray}
We see, that the current vertex is
proportional to $\left( 1 + |\omega| \tau
\right)^{-1}$ and its frequency dependence is determined by the two pairs
of $\theta$--functions in Eq.(A1). These terms exist, when the poles
$\varepsilon$ and $(\varepsilon - \omega)$ are located in the opposite
half-planes of the complex plane $\xi$. Let us note, that the similar situation
takes place when calculating Drude conductivity of the normal metal. 
Getting Eqs.(\ref{pi0}), (\ref{pi1}) and (\ref{vot}) together and using the following formula 
for the current response operator
\begin{equation}
\label{Qcl}
Q(\omega) =8 \nu e^2 T \sum\limits_{\Omega}
\gamma^2_{01}(\Omega,\omega) {\cal L}_0(\Omega) {\cal L}_1,
\end{equation}
we can calculate the AL contribution to the conductivity. Note, that in the framework of our
approximation, we can treat ${\cal L}_1$ as a constant, since it does not have any
linear $\Omega$-dependence (see Eq.(\ref{pi1})).  
The analytical continuation yields the following
expression  for the conductivity (valid within the logarithmic accuracy):
\begin{equation}
\delta\sigma = {\overline{\delta\sigma} \over \left(1 - i\omega \tau  \right)^2} =
{e^2 \over  \pi^2}
{1 \over \left(1 - i\omega \tau  \right)^2} 
\Bigl[ I_{\alpha}(h,t) + I_{\beta}(h,t) \Bigr],
\label{clcond}
\end{equation}
where functions $I_{\alpha}(h,t)$ and $I_{\beta}(h,t)$
are defined by Eqs.(\ref{Ia}, \ref{Ib}) with parameter
$r = \sqrt{\pi \over 4 \gamma}\, {h \over t}$, which
differs by a constant from the one in the dirty case.  

Equation (\ref{clcond}) is valid for $\omega \ll T_{c0}$. Recall, that the Drude conductivity has the following form:
\begin{equation}
\label{drude}
\sigma_0 = {\overline{\sigma_0} \over 1 - i \omega \tau} = {n e^2 \tau \over m} {1 \over 1 - i
\omega
\tau}.
\end{equation}
Thus, the total longitudinal resistivity reads
\begin{equation}
\label{res}
\rho_{xx} = {m \over n e^2} \left( {1 \over \tau} - i \omega \right) - {\overline{\delta\sigma}
\over \overline{\sigma_0}^2}.
\end{equation}
We see, that the fluctuation correction to the resistivity does not depend on the external frequency (unless $\omega \sim T_{c0}$)
and can be considered as a correction to the collision integral $1 \over \tau$. 
Physically, this means, that the AC electric field acts on the normal 
electrons, rather than superconducting fluctuations. 

In the superclean case $\omega_c \tau \lesssim 1$, we have to take into account the
curving of the classical trajectories. This curving results in the Hall term in the conductivity and
cyclotron resonance-like effects. The Hall term can be written as
$$
\rho_{xy} = {m \over n e^2}\, \omega_c  + \delta\rho_{xy},
$$ 
where the second term is due to the superconducting fluctuations.
The reasonings described above suggest that this term, which describes the curving of the
fluctuating pairs, is of the order of $\omega_c / T_{c0}$ and can be neglected.  
Hence, calculating inverse matrix $\hat\rho^{-1}$ we find the following formula 
for the fluctuation conductivity
\begin{equation}
\label{spm}
\delta\sigma_{\pm} = \delta\sigma_{xx} \pm i \delta\sigma_{xy} = 
{1 \over \left[ 1 - i \left( \omega \pm \omega_c \right) \tau \right]^2}  \overline{\delta\sigma},
\end{equation}
where $\overline{\delta\sigma}$ is defined by Eq.(\ref{clcond}). 
Let us note, that $\overline{\delta\sigma}$ represents the longitudinal conductivity
with no respect to the curving.
The corresponding Hall term $\overline{\delta\sigma_{xy}}$ 
can only appear in the presence of a particle-hole asymmetry. It does not exist in the framework of our
approximation. In the paper of Aronov {\em et al.} \cite{AHL} this additional Hall term was controlled by the phenomenological parameter 
$T_{c} {\partial \ln{T_c} \over \partial \varepsilon_F}$. 

Let us now discuss the contributions coming from the MT and DOS diagrams.
The electromagnetic response tensor can be written in the following
form:
\begin{equation}
\label{mtdos}
Q_{\alpha \beta}(\omega) = 2 e^2 T^2
\sum\limits_{\Omega,\, \varepsilon}
{\rm Tr\,} \left[ \hat B_{\alpha \beta} \left( \varepsilon, \omega, \Omega \right)
\hat{\cal L}(\Omega) \right],
\end{equation}
where $\hat B_{\alpha \beta}$ represents a four Green function block.
Let us consider this quantity on the example of the MT term.
In the coordinate representation within semiclassical approximation it
has the form:
\begin{eqnarray}
\label{Bexp}
 B_{\alpha \beta} \left( \varepsilon, \omega, \Omega;
{\bf r},{\bf r}' \right) &=&
\tilde B_{\alpha \beta} \left( \varepsilon, \omega, \Omega;
{\bf r} - {\bf r}' \right)
\nonumber \\
&\phantom{=}& \times \exp{ \left( - 2 i e \int\limits_{\bf r}^{\bf r'}{\bf A}({\bf s})
d{\bf s} \right)},
\end{eqnarray}
where 
\begin{eqnarray}
{\tilde B}_{\alpha \beta} \left( \varepsilon, \omega, \Omega;
{\bf r} \right) &=& 
- {r_{\alpha} r_{\beta} \over r^2}
\int\limits_0^{\infty}
{dp dk \over (2 \pi)^2}
J_1(pr) J_1(kr) \nonumber \\
&\times& {\cal G}_{\varepsilon}(p) {\cal G}_{\varepsilon-\omega}(p)
{\cal G}_{\Omega - \varepsilon}(k) {\cal G}_{\Omega + \omega -\varepsilon}(k).
\label{bgggg}
\end{eqnarray}
Putting this expression into the HW theorem (\ref{hwt}), evaluating the diagonal
matrix element for $n=0$ and performing integration over ${\bf r}$
we obtain the following expression
\begin{eqnarray}
\label{B00}
&\langle& 0 | \hat{B}_{xx}\left( \varepsilon, \omega, \Omega \right)| 0
\rangle = 
-\int\limits_0^{\infty}
{dpdk \over 2 \pi} p^2 k^2 \nonumber \\
\times &{\cal G}&_{\varepsilon}(p) {\cal G}_{\varepsilon-\omega}(p) 
{\cal G}_{\Omega - \varepsilon}(k) 
{\cal G}_{\Omega + \omega -\varepsilon}(k)
I_{00}(p,k),
\end{eqnarray}
where
\begin{equation}
\label{I00}
I_{00}(p,k) = r_H^2 \exp{\left[-r_H^2 \left(p^2 + k^2 \right) \right]}
I_1\left[ {1 \over 2} r_H^2 pk \right]
\end{equation}
and $I_1$ is the modified Bessel function of the first order.
Since $p \sim k \sim p_F$ and, thus, 
$r_H^2 pk \sim {\varepsilon / \omega_c} \gg 1$, 
we can take the asymtpotical form of the Bessel function and then
approximate the resulting exponent 
$\exp{\left[ - r_H^2 \left( p - k \right)^2 \right]}$ by
the $\delta$-function.
Thus, we have 
\begin{equation}
\label{I00del}
I_{00}(p,k) \approx {1 \over p} \delta(p-k).
\end{equation}
Performing integration with respect to ${\bf k}$ and introducing
the density of states at the Fermi-surface $N(0)$ we obtain
the usual expression for the 4-Green function block
\begin{eqnarray}
\label{finB} 
\langle &0& | B_{xx}\left( \varepsilon, \omega, \Omega \right)| 0
\rangle = - {1 \over 2} v_F^2 N(0) 
\nonumber \\
&\times& \int\limits_{-\infty}^{+\infty}
d \xi_{\bf p} 
{\cal G}_{\varepsilon}(p) {\cal G}_{\varepsilon-\omega}(p)
{\cal G}_{\Omega - \varepsilon}(p) {\cal G}_{\Omega + \omega -\varepsilon}(p).
\end{eqnarray}
One can see, that the derivation of this expression does not depend on the purity
of a superconductor. It is valid for the dirty and clean limits and any magnetic
fields applied, unless $\omega_c \sim \varepsilon_F$. 

Let us note that Eq.(\ref{finB}) for the MT diagram and the similar expressions for
the DOS diagrams are identical to the ones in the vicinity of $T_{c0}$ and
do not involve magnetic field at all. It is known, that DOS and MT terms are strongly compensated
in the clean limit \cite{Varlam} and this compensation takes place
at the level of Green functions ({\em i.e.} before integrals over ${\bf q}$). 
This suggests that, in the clean limit the only remaining diagram is the AL term 
even in the case of strong magnetic field.

Let us now discuss quantum oscillations in the fluctuation conductivity.
At very low temperature these oscillations 
become important. In this case, each current vertex contains an
oscillating part. This oscillating part can be found
by comparison with the Drude conductivity and can be written as
$\gamma = \gamma_0 + \gamma_{\rm osc}$, where 
$\gamma_{\rm osc} / \gamma_0 \sim \sigma_{\rm osc} / \sigma_0$ 
with $\sigma_{\rm osc}$ being the oscillating part of the normal conductivity
(see {\em e.g.} Ando {\em et al.} \cite{AFS})
However, there are other ``sources'' of quantum oscillations.
The transition temperature $T_{c}(B)$ oscillates as well \cite{GG,Tezan} and
affects the fluctuation conductivity.
Let us also realize, that 
the oscillations of magnetization (de Haas-van Alphen
oscillations) can influence Shubnikov-de Haas oscillations. Under certain
conditions, this effect may be dominant. 
Moreover, magnetization fluctuates as well and in the vicinity 
of the transition the fluctuations can exceed Landau 
diamagnetism (see section 3 and Ref. \onlinecite{AL2}). 
We see, that the oscillating part of the fluctuating
conductivity has a complicated structure and can differ significantly
from the usual Shubnikov-de Haas oscillations.

\section{Thermodynamics. Fluctuating magnetization}

\subsection{Dirty case}
\label{sec:2}

Considering thermodynamic properties of a film, we can calculate
the free energy directly. In the one-loop approximation, the free energy
can be written as \cite{Amb}
\begin{equation}
\label{f1}
F_1 = - T \sum\limits_{\Omega} {\rm Tr\,} \ln{
\left( 1 - g\, \hat C(\Omega) \right)},
\end{equation}
where $\hat C(\Omega)$ is the cooperon.

Using Eqs.(\ref{K}), (\ref{Coop}) and (\ref{f1}), one can easily obtain the magnetization
\begin{equation}
\label{M0}
M_1 = - {1 \over V} {\partial F_1 \over \partial H} =
{\nu \over 2 \pi d} {\Omega_H \over H_{c2}(0)}
I_{\alpha}(h,t),
\end{equation}
where $d$ is the thickness of the film or the interlayer distance,
$\nu={e H} /\pi$ is the number of states of a Landau level and
function $I_{\alpha}(h,t)$ is defined in Eq.(\ref{Ia}).
Thus, at low temperature $t \ll h$ the susceptibility takes the form:
\begin{equation}
\label{M0t0}
{\chi}_1 = - {\partial M_1 \over \partial H} ={e^2 \over \pi^2 \hbar c^2} {v_F^2 \tau \over d}\, h^{-1}. 
\end{equation}
One can see, that the fluctuation susceptibility (\ref{M0t0}) is large compared to the 
magnetic susceptibility 
of the normal metal $\chi_{\phantom{.}_{\rm L}}$  even far from the transition:
\begin{equation}
\label{chiL}
\chi_1 \sim {1 \over {\rm Gi}\, h}\, \chi_{\phantom{.}_{\rm L}},
\end{equation}
where ${\rm Gi}\, = \left( \varepsilon_F \tau \right)^{-1}$ is the Ginzburg parameter.

\subsection{Clean case}
\label{tc}

The calculation of magnetization in the clean limit can be done in the same fashion as in the dirty limit.
However, there are some features specific for the clean case. As we have already mentioned,
de Haas oscillations
become essential at low temperature in pure samples. These quantum oscillations appear in all
quantities including Green functions, transition temperature $T_c(H)$, fluctuating conductivity {\em etc.}
The oscillating terms are proportional to the factors $\exp{\left( -{\pi \over \omega_c \tau} \right)}$ and 
$\exp{ \left( - {2 \pi^2 T \over \omega_c} \right)}$. Hence, the oscillations are strongly suppressed, unless 
$\omega_c \tau \sim 1$ and $T / \omega_c \sim 1$. Let us note, that de Haas-van Alphen oscillations
in the magnetization can reveal themselves much earlier than the quantum oscillations in the other quantities.
This is because magnetization is a derivative of the free energy with respect to the magnetic field.
Even though the oscillating terms in the Green functions are small, they contain fast-oscillating
functions $\cos{ \left( 2 \pi \varepsilon_F / \omega_c \right)}$ which may lead to observable effects in
the oscillating magnetization. It is easy to calculate the fluctuating magnetization with respect
to these effects.

We can use Eq.(\ref{f1})
\begin{equation}
\label{f1cl}
F_1 = - T \sum\limits_{\Omega} {\rm Tr\,} \ln{
\left( 1 - g\, \hat \Pi(\Omega) \right)},
\end{equation}
where $\hat \Pi = \hat \Pi_0 + \hat \Pi_{\rm osc}$ is the particle-particle bubble which contains
an oscillating part. The matrix element for the monotonous part of the particle-particle bubble corresponding to the lowest Landau level 
was found in Sec. \ref{sec:3} (see Eq.(\ref{pi0})). The oscillating part 
has been considered in a number of papers and has the form: \cite{GG,Min}
\begin{eqnarray}
\label{s1}
\Pi_{\rm osc} &=& - 8 \pi^{3/2} N(0) {T \over \sqrt{\varepsilon_F \omega_c}}\,
\cos{ \left( 2 \pi {\varepsilon_F \over \omega_c} \right)}
\nonumber \\
&\times&
\cos{ \left( 6 \pi {\mu_e H \over \omega_c} \right)} 
\exp{\left(- {\Delta \over \omega_c } \right)},
\end{eqnarray} 
where $\Delta = 6 \pi \left(  \pi T + {1 \over 2 \tau} \right)$ and $\mu_e$ is 
the magnetic moment of an electron.
For the sake of simplicity, we keep the first oscillating term only.

In the vicinity of the transition we can present the magnetization in the following way 
\begin{equation}
\label{Mosc}
M_1 = {T \nu \over H_{c2}(0)}\, 
\sum\limits_{\Omega}
{1 \over {\cal L}_0(\Omega)^{-1} - \Pi_{\rm osc}}
{\partial \over \partial h}
\left( {\cal L}_0(\Omega)^{-1} - \Pi_{\rm osc} \right),
\end{equation}
where for ${\cal L}_0(\Omega)$ see Eq.(\ref{l0c}).

From Eqs.(\ref{l0c}), (\ref{f1}) and (\ref{s1}), we obtain the following
expression for the fluctuating magnetization
\begin{eqnarray}
\label{resmag}
M_1 &=& {1 \over \sqrt{\pi \gamma}} \, { T_{c0} \nu \over H_{c2}(0)}
\left[ \ln{1 \over t} - \psi\left( {1 \over \sqrt{4 \pi \gamma}} {h \over t} \right) -
 \sqrt{\pi \gamma}\, {t \over h} \right]
\nonumber \\
&\phantom{=}& \times \Biggl[ \Biggr. 1 + 32 \pi^{5/2} {T \sqrt{\varepsilon_F} \over
\omega_c^{3/2}}  
\sin{\left( 2 \pi {\varepsilon_F \over \omega_c} \right)}
\nonumber \\
&\phantom{=}& \times
\cos{ \left( 6 \pi {\mu_e H \over \omega_c} \right)} \exp{\left( - {\Delta \over \omega_c} \right)}
\Biggl. \Biggr].
\end{eqnarray}
Let us note, that if $T \sim \omega_c$, than  $T \sqrt{\varepsilon_F} / \omega_c^{3/2} \sim
{\varepsilon_F / T_{c0}}
\gg 1$ and the numerical factor in the oscillating term in Eq.(\ref{resmag}) is very large. 
Thus, we conclude that de Haas-van Alphen oscillations in magnetization may exist even 
in the absence  of the Shubnikov-de Haas oscillations and oscillations of the transition
temperature.  It is worth mentioning, that the fluctuation effects exceed Landau diamagnetism in the
clean limit as well (formula (\ref{chiL}) is valid with ${\rm Gi\,} = \varepsilon_F / T_{c0}$).  
Thus, under certain circumstances ($\Delta \sim 1$) the oscillating part of the fluctuating
magnetization may be more important than the monotonous part of magnetization and the oscillating
part in the Landau term.

\section{Two-loop approximation. Applicability of the results.}
\label{2l}

In the previous sections we found the fluctuation correction to the the
transport and thermodynamic properties of a superconductor in a magnetic field in the
first (one-loop) approximation.
The purpose of the given section is to find the order of the subleading 
corrections. This will determine
the area of applicability of the results obtained.
We shall calculate the magnetization in the two-loop approximation
for a dirty superconductor. 
This correction can be easily calculated in view of the simplifications
described above.
 
In the two-loop approximation we have to deal with diagrams presented on Fig. 4. The corresponding 
contribution can be written in the coordinate representation in the following way
\begin{eqnarray}
\label{f2}
F_2 &=& T^3 \sum\limits_{\varepsilon, \Omega, \Omega'}
\int d^2 r_1 d^2 r_2 d^2 r_3 d^2 r_4
\nonumber \\
&\phantom{=}& \times K_{\varepsilon} ({\bf r}_1,{\bf r}_2;{\bf r}_3,{\bf r}_4)
{\cal L}_{\Omega}({\bf r}_1,{\bf r}_2)  
{\cal L}_{\Omega'}({\bf r}_3,{\bf r}_4),
\end{eqnarray}
where $K_{\varepsilon}$ is the operator corresponding to the square blocks in the diagrams presented on Fig. 4.
This operator is familiar from the usual BCS theory. It has been calculated by Maki \cite{Mak2} and Caroli {\em et al.} \cite{deG} and has the  form:
\begin{eqnarray}
&K_{\varepsilon}& ({\bf r}_1,{\bf r}_2;{\bf r}_3,{\bf r}_4) =
{\pi N(0) \over 2} \nonumber \\
&\times&
\delta({\bf r}_1 - {\bf r}_2) \delta({\bf r}_1 - {\bf r}_3) \delta({\bf r}_1 - {\bf r}_4) 
\left\{ \prod\limits_{k=1}^4
{1 \over |\varepsilon| +{1 \over 2} {\cal D} \partial_{(k)}^2 }\right\}
\nonumber \\
&\times& \Biggl[ \Biggr. |\varepsilon| + {1 \over 8} {\cal D}
\Bigl( \Bigr. \left[ \partial_{(1)} - \partial_{(3)} \right]^2 
+
\left[ \partial_{(2)} - \partial_{(4)} \right]^2 \Bigl. \Bigr) \Biggl. \Biggr],
\end{eqnarray}
where we make use of the Maki's notations:

\narrowtext
\begin{figure}
\epsfxsize=8.5cm
\centerline{\epsfbox{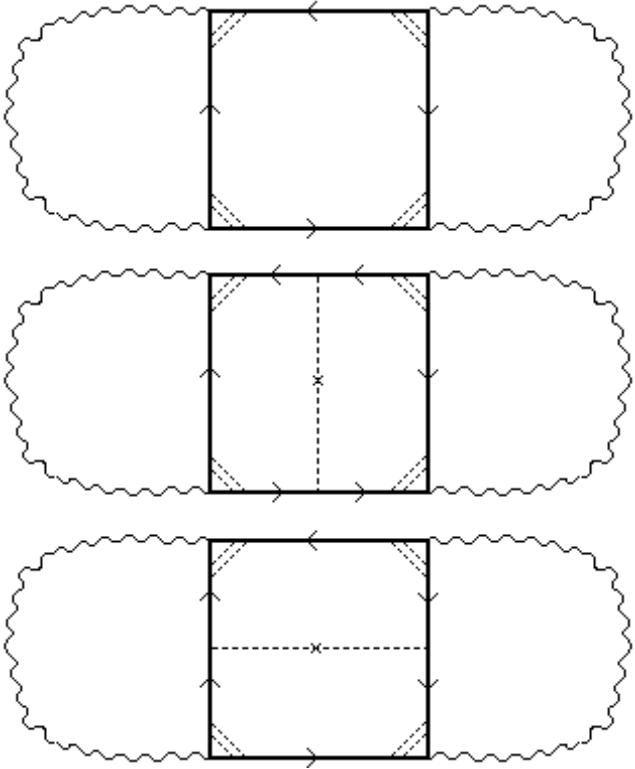}}
\caption{\label{fig:K}
Diagrams contributing to the free energy in the two-loop approximation.
Similar diagrams appear in the derivation of the Ginzburg-Landau equations
from the microscopic theory. 
}
\end{figure}

$$
\partial_{(k)} = -i {\bf \nabla} - 2 e (-1)^k {\bf A}({\bf r}).
$$
In the coordinate representation, the fluctuation propagator can be
expanded on the basis of the eigenfunctions in the magnetic field
and has the form: 
\begin{equation}
\label{Lpp}
{\cal L}_{\Omega} ({\bf r},{\bf r}') =
\int\limits_{-\infty}^{+\infty} {d{p_y} \over 2 \pi} \sum\limits_{n=0}^{\infty}
{\cal L}_n(\Omega) \psi_{n{p_y}}^* ({\bf r}) \psi_{n {p_y}}({\bf r}'),
\end{equation}
where ${\cal L}_n(\Omega)$ are matrix elements of the fluctuation propagator
in the magnetic field (see Eq.(\ref{K})),
$\psi_{n {p_y}}({\bf r}')$ is the eigenfunction for an electron in
a magnetic field in the Landau gauge and $p_y$ is the $y$-component of the
momentum, which determines the orbit's center.
Again, in the vicinity of the transition line we keep the $n=0$ term only
in Eq.(\ref{Lpp}). From Eqs.(\ref{f2}---\ref{Lpp}), we obtain
the free energy per unit volume
\begin{equation}
\label{Fa}
{F_2 \over V} =
{\pi N(0) \over 2 d} \nu^2 T^3
\left( \sum\limits_{\Omega} L_0(\Omega) \right)^2
\, \sum\limits_{\varepsilon}
{1 \over \left( \left| \varepsilon \right| +\Omega_H/2 \right)^3}.
\end{equation}
Thus, the magnetization takes the form:
\begin{equation}
\label{M1}
M_2 = {\nu^2 \over (2 \pi)^2 d N(0)} \,
{1 \over H_{c2}(0)} 
{\partial I_{\alpha}^2(h,t) \over \partial h}.
\end{equation}
At low temperatures $t \ll h$ we have
\begin{equation}
\label{M1t0}
M_2 = -{\nu^2 \over 2 \pi^2 d N(0)} {1 \over H_{c2}(0)}
{1 \over h} \ln{1 \over h}.
\end{equation}
We see, that the second order correction is negative. 

From Eqs.(\ref{M0}) and (\ref{M1}) we obtain the ratio
\begin{equation}
\label{ratio}
{M_2 \over M_1} = {{\rm Gi} \over 2 \pi}\,
\left[ \gamma {t \over h^2} - {1 \over 2 \gamma t} \psi'\left( {1 \over 2 \gamma} {h \over t}
\right) \right],
\end{equation}
where ${\rm Gi}$ is the Ginzburg parameter. The one-loop approximation is valid unless this ratio becomes of the order of unity.
At low temperatures $t \ll h$, Eq.(\ref{ratio}) yields the following condition
\begin{equation}
\label{Gi1}
h \gg {\rm Gi}.
\end{equation}
If $t \gg h$, we have
\begin{equation}
\label{Gi2}
h \gg \sqrt{ {\rm Gi\,} t}.
\end{equation}
This indicates that at large enough temperatures the fluctuation region becomes wider. 

These results stand for the kinetic coefficients as well.
In the clean case the formulae (\ref{Gi1}) and (\ref{Gi2}) are valid with
${\rm Gi} \sim {\varepsilon_F / T_{c0}}$. However, explicit calculations are more complicated
due to the non-local structure of the $K$-operator.  

Let us note, that at an exponentially low temperature some other effects may reveal themselves.
In the dirty case, mesoscopic fluctuations may be important. \cite{Spiv}
Really, the upper critical field depends on disorder. The distribution of impurities
is random. There are some regions where the concentration of the impurities
is such that the upper critical field is smaller than the bulk value. These regions
may form superconducting islands weakly coupled one with another. At extremely
low temperature the proximity effect and the Josephson coupling can
make these mesoscopic fluctuations observable. The effects due to the mesoscopic
fluctuations will be considered elsewhere.

\section{Conclusion}

The central result of the paper is the existence of the logarithmic
correction to the conductivity which persists down to zero temperature.
This correction is shown to be negative in the dirty case. The minus sign
comes from the DOS diagrams as well as from the
anomalous MT term. The AL contribution is positive but numerically smaller. 
Let us note, that similar results (negative fluctuation
correction to conductivity) exist for the granular and layered superconductors. \cite{BEL,Varobzor}
In these cases the AL and MT contributions are parametrically small
compared to the DOS term. 

The fluctuating magnetization exceeds conventional Landau diamagnetism
for a very large range of fields. It is shown to be logarithmically 
divergent as well at $T \to 0$. 

Let us note, that singular behavior of transport and thermodynamic
quantities at low temperature is due to the low dimensionality
of the system. 
In the three dimensional case the leading correction to conductivity
is not singular $\delta \sigma_{\rm 3D} \sim \sqrt{h}$.

The results obtained in the present paper can be checked experimentally
by measuring fluctuation conductivity in two-dimensional and 
quasi-two-dimensional systems.  The results obtained
in the dirty limit can be checked by measuring magnetoresistance
in dirty superconducting films at low temperature. In this case, there could be some experimental difficulties
connected with the $H_{c3}$-effects that can screen the bulk properties of
a film. The edge effects can be excluded, for example, by putting a sufficient
amount of magnetic impurities on the edge of the film.

The clean case may be re\-le\-vant to 
high-Tc su\-per\-con\-duc\-tors \cite{Boeb} and, probably, to the re\-ce\-ntly dis\-co\-ve\-red
two-dimensional organic superconductors. \cite{Bat} Let us note, that our results
assume s-pairing and isotropic Fermi-surface which is not true
for High-$T_c$ superconductors. However, it can be shown, that the logarithmic singularity
$\ln{1 \over h}$ remains for any pairing type (with a coefficient
different from our case). It is worth mentioning, that in the 
overdoped High-$T_c$ superconductors the Ginzburg parameter ${\rm Gi}$
is small and, thus, the fluctuations are negligible. In the underdoped
superconductors the fluctuations are extremely large and they lead to a
large pseudogap which makes the conventional Fermi-liquid theory inapplicable. 
Hence, optimally doped superconductors should be used to check the results obtained.

\acknowledgments
V. Galitski wishes to thank Alexei Kaminski for all his help.
This work was supported  by NSF grant DMR-9812340.

\appendix
\section{Current vertex}

\label{sec:a}

In this Appendix we derive the formula for the current vertex depending
on two frequencies $\Omega$ and $\omega$ in the non-local clean limit.
The corresponding result is used when calculating the Aslamazov-Larkin
contribution to the conductivity (see Sec. \ref{sec:3}).

The current vertex is the triangle block in the AL diagram (see Fig. 1.1).
It consists of three Green functions.
In the momentum representation, it can be written as

\begin{eqnarray}
\label{c}
{\bf \gamma}_{\varepsilon} ({\bf q}; \Omega, \omega) 
&=& \int
{d^2 p \over (2 \pi)^2}
{\bf v}\, 
G_{\varepsilon}({\bf p})
G_{\varepsilon-\omega}({\bf p}) 
G_{\Omega-\varepsilon}({\bf q} - {\bf p}) \nonumber \\
&=& - N(0) \int d\xi {1 \over \xi - i \tilde\varepsilon}  
{1 \over \xi - i \left( \widetilde{\varepsilon - \omega} \right) }
\nonumber \\
&\phantom{=}& \times \left\langle{{\bf v} \over \xi - {\bf vq} - i \left( \widetilde{\Omega - \varepsilon} \right) } \right\rangle,
\end{eqnarray}
where $\tilde\varepsilon = \varepsilon + {i \over 2 \tau} {\rm sgn \,}\varepsilon$ and 
the angular brackets imply averaging over the Fermi-line.
To perform this averaging one can use the following identity:
\begin{equation}
\label{ugol}
\left\langle {{\bf v} \over {\bf vq} - i \epsilon} \right\rangle
={{\bf q} \over q^2} \left( 1 - 
{\left| \epsilon \right| \over \sqrt{\epsilon^2 + v_F^2 q^2}} \right) \equiv {{\bf q} \over q^2} f(\epsilon, q).
\end{equation}

There are six possible configurations of the poles which give 
non-zero contributions to the integral over $\xi$ in Eq.(\ref{c}). Straightforward
calculation yields the following expression for the current vertex
\begin{equation}
\label{gama}
{\bbox \gamma}_{\varepsilon} ({\bf q}; \Omega, \omega)
= {2 \pi N(0) \over \omega} \, {{\bf q} \over q^2} \,  
\sum\limits_{i=1}^4 \eta_i(\varepsilon,\Omega,\omega) f(\epsilon_i,q) \, ,
\end{equation}
where
\begin{eqnarray}
\label{th1}
\eta_1(\varepsilon,\Omega,\omega) = &\phantom{.}& {\omega \tau \over 1 + \omega \tau\, {\rm sgn \,} \varepsilon} 
\Bigl( \theta(\varepsilon) \theta(\omega - \varepsilon) \theta (\varepsilon - \Omega) 
\nonumber \\ &+&
\theta(-\varepsilon) \theta(\varepsilon-\omega) \theta (\Omega - \varepsilon) \Bigr),
\end{eqnarray}
\begin{eqnarray}
\label{th2}
\eta_2(\varepsilon,\Omega,\omega) = &\phantom{.}& {\omega \tau \over 1 + \omega \tau\, {\rm sgn \,} \varepsilon} \,
\Bigl( \theta(\varepsilon) \theta(\omega - \varepsilon) \theta (\Omega -\varepsilon) 
\nonumber \\ &+&
\theta(-\varepsilon) \theta(\varepsilon-\omega) \theta (\varepsilon - \Omega) \Bigr),
\end{eqnarray}
\begin{eqnarray}
\label{th3}
\eta_3(\varepsilon,\Omega,\omega) = &-& \eta_4 (\varepsilon,\Omega,\omega) = 
\Bigl( \theta(\varepsilon) \theta(\varepsilon - \omega) \theta (\varepsilon - \Omega) 
\nonumber \\ &-&
\theta(-\varepsilon) \theta(\omega - \varepsilon) \theta (\Omega - \varepsilon) \Bigr)
\end{eqnarray}
and
\begin{equation}
\label{e14}
\epsilon_1 = \epsilon_4 = 2 \varepsilon - \Omega +  \tau^{-1}\, {\rm sgn \,} \varepsilon,
\end{equation}
\begin{equation}
\label{e23}
\epsilon_2 = \epsilon_3 = 2 \varepsilon - \Omega - \omega +  \tau^{-1}\, {\rm sgn \,} \varepsilon.
\end{equation}
This presentation is convenient when we calculate the current vertex in the magnetic field (Sec. {\ref{sec:3}}).

In the dirty limit Eq.(\ref{gama}) reduces to the local equation, since in the limit $\tau \to 0$
$$
f(\epsilon, q) \approx {\cal D} q^2 \tau
$$
and thus the current vertex reads 
\begin{equation}
\label{gamaloc}
{\bbox \gamma}({\bf q})= c \, {\bf q} \equiv - 4 \pi N(0) {\cal D} \tau^2 {\bf q}.
\end{equation}
Here we keep only the terms which do not contradict to the theta-functions in the cooperons, {\em i.e.} 
the third and fourth terms in Eq.(\ref{gama}).
We see, that in the coordinate representation the vertex has the form:
$$
{\bbox \gamma} ({\bf r}) = -c {\bf \nabla} \delta({\bf r}).
$$
This $\delta$-functional behavior implies the locality of the current vertex in the dirty case.

Let us also note, that if the external frequency $\omega$ is zero, the current vertex is easily connected
with the particle-particle bubble for any $\tau$:
\begin{equation}
\label{dP}
\lim\limits_{\omega \to 0} {\bbox \gamma}_{\varepsilon}({\bf q}; \Omega, \omega) 
= {\partial \Pi_{\varepsilon} ({\bf q}; \Omega) \over \partial {\bf q}},
\end{equation}
where
\begin{eqnarray}
\label{bubble}
\Pi_{\varepsilon} ({\bf q}; \Omega) &=&
N(0) \left\langle
\int d \xi {\cal G}_{\varepsilon}({\bf p})  
{\cal G}_{\Omega - \varepsilon}({\bf q} - {\bf p}) \right\rangle 
\nonumber \\
&=&
2 \pi N(0)
{ \theta \left(\varepsilon \left( \varepsilon - \Omega \right) \right) \over
\sqrt{
\left( 2 \varepsilon - \Omega + {1 \over \tau} {\rm sgn \,}\varepsilon \right)^2 + v_F^2 q^2}}.
\end{eqnarray}

\end{multicols}

\end{document}